\documentclass[runningheads]{llncs}

\pdfoutput=1
 
\usepackage[utf8]{inputenc}
\usepackage[russian,english]{babel}
\usepackage{array}
\usepackage{makeidx} 
\usepackage{marvosym}
\usepackage{amsfonts}
\usepackage{amsmath}
\usepackage{mathrsfs}
\usepackage{graphicx}
\usepackage{url}
\usepackage{hyperref}
\usepackage{wasysym}
\usepackage{caption}
\usepackage{subcaption}

\usepackage{xcolor}

\newcolumntype{L}[1]{>{\raggedright\let\newline\\\arraybackslash\hspace{0pt}}m{#1}}
\newcolumntype{C}[1]{>{\centering\let\newline\\\arraybackslash\hspace{0pt}}m{#1}}
\newcolumntype{R}[1]{>{\raggedleft\let\newline\\\arraybackslash\hspace{0pt}}m{#1}}

\begin{document}

\title{Personal Names Popularity Estimation \\and its Application to Record Linkage\thanks{Work carried out while authors at Kontur Labs, a research department of SKB Kontur, \url{https://kontur.ru/eng/}. This is an extended version of a short paper presented at ADBIS2018.}}

\titlerunning{Personal Names Popularity Estimation}

\author{Ksenia Zhagorina \inst{1} \and Pavel Braslavski\inst{2} \and Vladimir Gusev\inst{2}}

\institute{Yandex, Yekaterinburg \email{Ksenia.Zhagorina@yandex.ru} \and Ural Federal University, Yekaterinburg \email{\{Pavel.Braslavsky,Vladimir.Gusev\}@urfu.ru}
}
\authorrunning{K. Zhagorina, P. Braslavski, V. Gusev}

\maketitle
\begin{abstract}

This study deals with a fairly simply formulated problem~-- how to estimate the number of people bearing the same full name in a large population. Estimation of name popularity can leverage personal name matching in databases and be of interest for many other domains. 
A distinctive feature of large collections of names is that they contain a large number of unique items, which is challenging for statistical modeling. We investigate a number of statistical techniques and also propose a simple yet effective method aimed at obtaining more accurate count estimates. In our experiments we use a dataset containing about 20 million name occurrences that correspond to about 13 million real-world persons. We perform a thorough evaluation of the name count estimation methods and a record linkage experiment guided by name popularity estimates. Obtained results suggest that theoretically informed approaches outperform simple heuristics and can be useful in a variety of applications.


\end{abstract}

\section{Introduction}

Our study deals with a fairly simply formulated problem~-- given a full name, how to estimate the number of people bearing this particular name in a large population? Originally, the study was motivated by an applied record linkage task in a large database, where occurrences of personal names were accompanied with no or only scarce additional information. 

Record linkage -- the task of matching records referring to the same real-world entity -- is a well-studied field within database technology that is also known under such names as \textit{name matching}, \textit{entity resolution}, \textit{object identification}, \textit{deduplication}, and others. 
The task arises when several databases are merged or one is interested in linking duplicate records within a single database. Records referring to people are the most common objects of linkage task. There is a wide variety of application domains, such as social network profiles, medical and census data processing, human resources and customer relationship management, bibliographic and genealogy databases, etc. Discrepancy in data can occur due to different attribute sets, as well as on individual fields' level due to misspellings and OCR errors, name (cf. nicknames) and transliteration variations. 
In contrast, in our settings there is no additional information but \textit{identical} names. Name popularity estimates can serve as an additional signal for matching in case of limited information.
In our study we use a large dataset of open government data that gave rise to the applied task initially. The dataset contains about 20 million records that correspond to 13.4 million real-word persons, which constitutes about one tenth of the entire Russian population. Russia is a multi-ethnic country and we may hope that methods described herein are not heavily dependent on language and culture and can be applied to other name collections. 

Accurate name popularity estimation based on limited number of observations is a hard task. Even very large collections contain many unique names -- names are a good example of \textit{large number of rare events (LNRE)} distributions. It is quite natural -- names serve primarily to distinguish people and to avoid collisions. Therefore maximum likelihood estimates based even on large name samples are poor predictors, since there are always many unseen names. A simplistic assumption that all names are unique may work in small communities, but in larger populations one can observe a whole spectrum from singletons to very common names.
To address the problem we employ several smoothing techniques that redistribute probability mass from already seen names towards yet unseen ones. In addition, we use the fact that first, middle and last names are dependent on each other. We use LNRE models to estimate the number of unique names and use this estimate as smoothing parameter. In case of full name triples (first, middle, and last name) we apply Markov assumption, i.e use only pairwise conditional probabilities. We also propose our simple yet effective technique for name count estimates that takes into account the large number of unique names. 

We conducted two experiments: 1) name popularity estimation and 2) record linkage guided solely by the name popularity estimates. We performed evaluation both for name triples (first, middle, and last) and doubles (first and last). Obtained results suggest that theoretically informed approaches outperform simple heuristics. Name popularity estimates can be a good supplemental signal in record linkage tasks, help distinguish unrealistic (artificial), rare and more common names. The main contribution of our study is a thorough comparative evaluation of several statistical techniques applied to the name popularity estimation task on a sizable dataset.
The study provides guidance for choosing the most appropriate model depending on available data, task, and performance requirements.
 
Knowing an estimate of people bearing a particular name is beneficial not only for record linkage in databases, but also for social network analysis (especially in detecting fake and duplicate accounts\footnote{Facebook estimated that duplicate and fraudulent accounts represented up to 14\% of its worldwide monthly active users in the fourth quarter of 2017, see \url{https://investor.fb.com/financials/sec-filings-details/default.aspx?FilingId=12512043}}), people search, information security, and information extraction. Quantitative analysis of personal names is also of interest for genealogy, demographic, sociological, and human biology research. Last but not least, name popularity estimates can be helpful in such an important matter as the choice of the name for a newborn baby.

The paper is organized as follows. Section~\ref{sec:relwork} gives an overview  of various studies on numerical name analysis; section~\ref{sec:data} describes the data used in the study. In section~\ref{sec:methods} we describe name popularity prediction models investigated within the experiment, as well as experimental design, evaluation approaches and measures. Section~\ref{sec:results} reports experimental results. Section~\ref{sec:conclusion} concludes and outlines directions of future work.

\section{Related Work}
\label{sec:relwork}

Our study is close to personal name matching~\cite{christen2006}, a special case of \textit{record linkage}~-- the task of matching records referring to the same real-world person in the presence of errors, spelling variants, omissions, abbreviation, etc. Most name matching method rely on pre-defined or machine-learned similarity measures for field values and tuples, see~\cite{christen2012,Ilyas2015,Naumann2010}. The main difference of our study is that we deal with \textit{identical} names and no additional fields. Our approach is close to record linkage methods that use conditional probabilities for field values (see, for example, an early work by Winkler~\cite{winkler1988}). However, we do not adjust our methods to a particular database; we rather aim at modeling name popularity at a global scale. As such, name popularity models can deliver additional evidence for record linkage tasks applied to different databases and in case of scarce additional information.
Personal name matching and deduplication attract a great deal of attention. For example, 238 teams participated in the Author Disambiguation Challenge in 2013.\footnote{\url{https://www.kaggle.com/c/kdd-cup-2013-author-disambiguation/}} The task was to identify which authors in a large bibliographic database correspond to the same person. The winning solution \cite{chin2014} used string similarity measures and an ensemble classifier for two concurrent matcher implementations, as well as processed Chinese and non-Chinese names separately. 
A recent Multilingual Web Person Name Disambiguation shared task\footnote{\url{http://nlp.uned.es/IberEval-2017/index.php/Tasks/M-WePNaD}} consisted of clustering Web search results for a person name query accounting for different real-world  persons~\cite{IberEval}.

In a related study Popescu et al. \cite{popescu2012} address the problem of estimating the number of people with identical names mentioned in a corpus in the context of information extraction. 
Authors assess name popularity  on the basis of phone books, Web search statistics, and name counts in Wikipedia. In contrast to our study, no evaluation of name popularity estimates was performed, as well as no formal justification of the method.

Name frequencies and their dynamics along with demographic information
can provide valuable insights for psychology~\cite{savage1948,zweigenhaft1980}, human biology~\cite{colantonio2003}, sociology and history~\cite{Scapoli2007,mateos2011}. The advent and proliferation of online social networks had a powerful impact on quantitative research on names, as name is often the only available information about the user. There is a series of studies that derive ethnicity~\cite{chang2010,mislove2011,rao2011} and gender~\cite{Bergsma2013,panchenko2014} from names in social network profiles. Perito et al.~\cite{Perito2011} and Liu et al.~\cite{liu2013} introduce the problem of linking user profiles belonging to the same physical person between online social networks based solely on usernames. These studies are relevant to ours since the central notion in both approaches is username uniqueness. The latter study models username unexpectedness with character-level Markov model. The authors of the former study first perform username segmentation; then estimate rareness or commonness of a segmented username using web n-gram statistics. Minkus et al.~\cite{minkus2015city} 
match population registry entries from a small US city to Facebook accounts based on straightforward name and location matching. Thonas et al.~\cite{fake_accounts} analyze naming patterns in fraudulent Twitter accounts.

Smoothing techniques we employ in the study have been actively developed within statistical language modeling~\cite{goodman2001,chen1999}.
Khmaladze \cite{khmaladze1988} introduced the notion of \textit{large number of rare events (LNRE)} distributions and studied their statistical properties. Baayen~\cite{baayen} and Evert~\cite{evert2004} elaborated the models for a better fitting of frequency distributions of words in large corpora, with special attention to the estimation of \textit{hapax legomena} count (that is, count of words with frequency 1). We use LNRE models for a more accurate choice of smoothing parameters in several evaluated methods. To the best of our knowledge, application of smoothing and LNRE models to the name popularity prediction task is novel. 

\section{Data}
\label{sec:data}

In our study we experiment with a dataset that originates from the \textit{Russian registry of legal entities and individual entrepreneurs}.\footnote{\url{http://egrul.nalog.ru/}} There is a many-to-many relationship between persons and companies: each legal entity is associated with one or more persons -- managers and/or founders; each real-word person can be associated with several companies. The registry contains about~32 million name mentions. Minimal piece of information about a person is his or her full name. Full names in Russian official documents are triples comprising of first, middle (patronymic), and last names, for example, \textit{Alexander Sergeyevich Pushkin}. Patronymics have gender-specific endings (cf. \textit{Sergeyevich} and \textit{Sergeyevna} -- literally Sergey's son and daughter, respectively) as many (but not all) Slavic last names do (\textit{Pushkin} and \textit{Pushkina} for male and female variants of the same family name, respectively). In our experiment we unify gender-specific variants of last names and patronymics. 

A subset of records contains persons' taxpayer identification numbers (TINs) that can be used as a key. In the rest of the paper we focus on about 20.6 million records containing both TIN and full name that refer to about 13.4 million real persons, which constitutes about one tenth of the entire Russian population.\footnote{According to the 2010 census, Russian population is 143,666,931, see \url{http://www.gks.ru/free_doc/new_site/perepis2010/croc/perepis_itogi1612.htm} (in Russian).} There are about 63.2 million pairs of identical names among 20.6 million occurrences, i.e. potential links between same-person records; 32\% of them are correct according to TINs. 

\begin{figure}[t]
        \centering
        \begin{subfigure}{0.5\textwidth}
                \includegraphics[width=\textwidth]{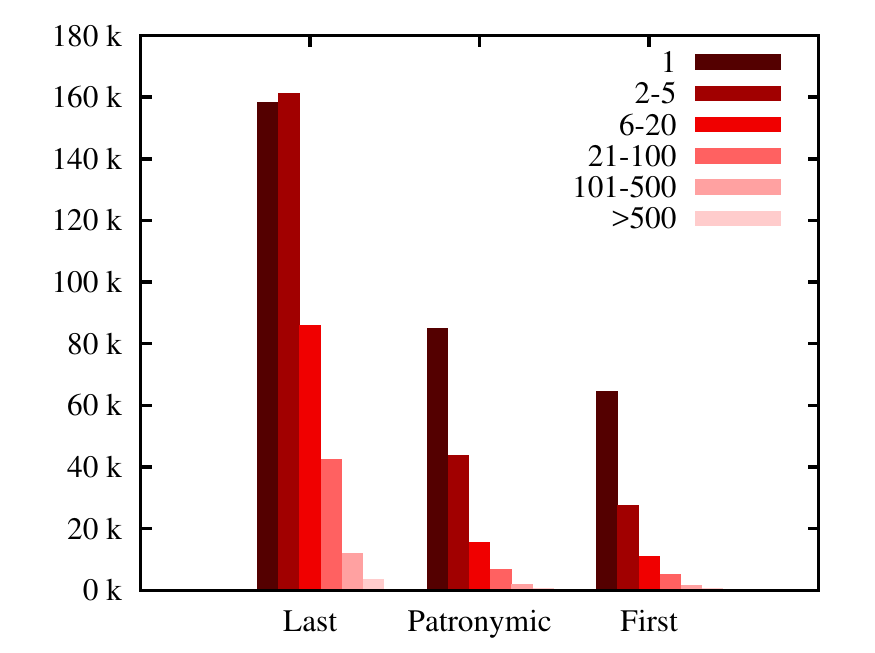}
                \caption{}
                \label{fig:names}
        \end{subfigure}%
        ~ 
        \begin{subfigure}{0.5\textwidth}
                \includegraphics[width=\textwidth]{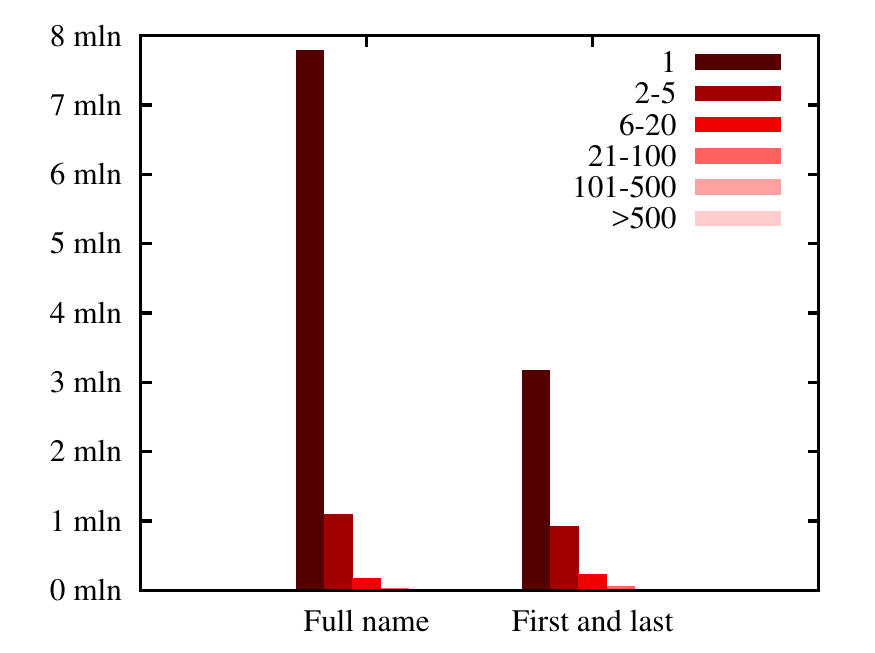}
                \caption{}
                \label{fig:names_comb}
        \end{subfigure}
        \caption{Histograms of name frequencies}
        \label{fig:namesComb}
\end{figure}

\begin{figure}[h]
\centering
\includegraphics[width=0.8\linewidth]{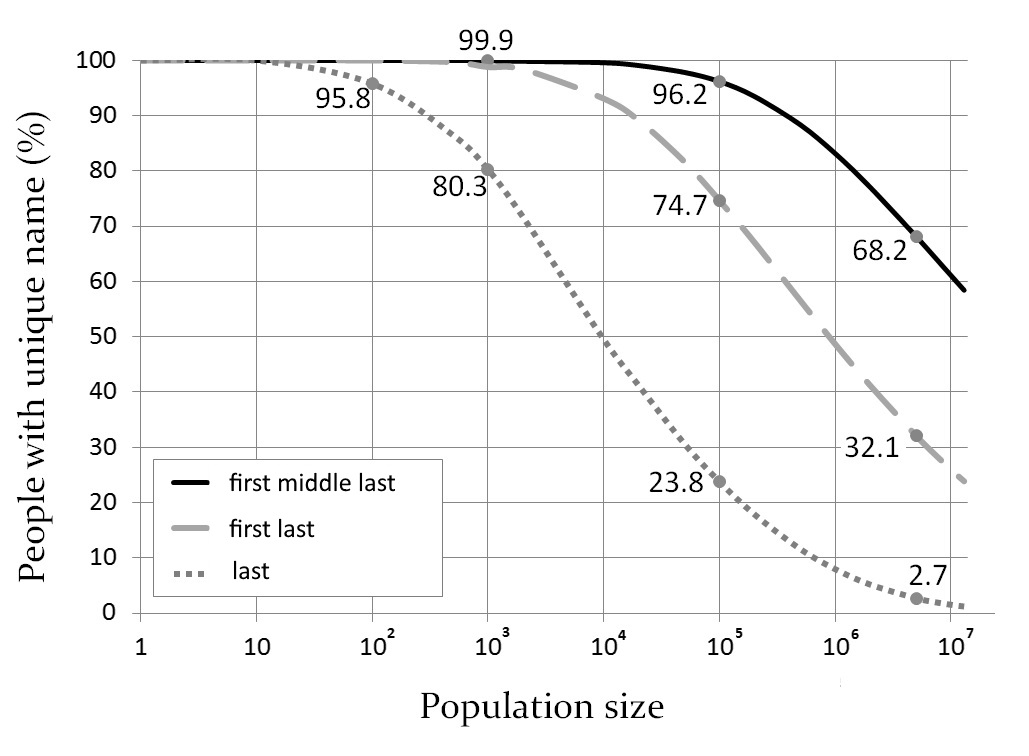}
\caption{Share of unique names depending on population size}
\label{fig:unique}
\end{figure}

Figure~\ref{fig:namesComb} illustrates that first, middle, and last names taken separately or as full names are a good example of \textit{LNRE} regime: the majority of names occur only once, while a small number of combinations are relatively common.
Expectedly, last names tend to be more rare than first names and patronymics (the latter are derivatives from male first names). Figure~\ref{fig:unique} shows proportions of unique name combinations in random samples of different sizes. For example, in a random population of 100,000 a combination of first, middle and last name is an almost perfect identifier (about 96\% people bear a unique name), while name pairs (first, last) reliably distinguish less then 75\% of people in the same sample.\footnote{Names of inhabitants of a particular city/region are presumably less diverse due to a higher ethnic and cultural homogeneity.} 

\section{Methods} \label{sec:methods}

\subsection{Name Popularity Prediction Methods}
\label{sec:namemodels}

In this section, we informally describe name popularity prediction models evaluated within the study. 
In what follows, $C(x)$ is the number of people with a name $x$ in training set $S_{train}$, where $x$ can be either a full name or its constituents;  $f$ stands for first name, $m$ and $\ell$  -- for middle and last names, respectively; $N_r$ is the number of names that occur exactly $r$ times in $S_{train}$ and $N$ is the total number of persons in $S_{train}$. 

We start with a na\"{i}ve approach assuming all people have unique names (model I). So, the number of people with the name $x$ is equal to 1 in the population of any size. Then, we proceed with straightforward maximum likelihood estimates (MLE) for full names (II):

\begin{equation}
\label{eq:mle}
P_{MLE}(fm\ell) = \frac{C(fm\ell)}{N} 
\end{equation}

Model II assigns zero probabilities to names unseen in the training set. To partially mitigate the problem we can assume independence of name constituents and approximate the probability of a full name by the product of individual first, middle, and last name probabilities (or just first and last name probabilities in case of name doubles), which defines model III: 

\begin{equation}
\label{eq:mult}
P_{ind}(fm\ell) = P_{MLE}(f) P_{MLE}(m) P_{MLE}(\ell) = \frac{C(f)}{N}\cdot \frac{C(m)}{N} \cdot \frac{C(\ell)}{N}
\end{equation}

\noindent This model assigns a zero probability to a name if one of its components is new in the test set.

Some combinations of first, middle, and last names occur together more frequently than others. The reasons are diverse: cultural and ethnic traditions, fashion (e.g. celebrities' names), or euphony of a combination. To capture these dependencies we use conditional probabilities. In case of names triples we apply Markov assumption, in other words -- we account only for dependencies between pairs of constituents\footnote{This approach corresponds to the bigram language model, however in case of names the order of constituents is irrelevant and we can experiment with different dependencies.}:
\begin{equation}
\label{eq:joint}
P(fm\ell) = P(f)P(m|f)P(\ell|f,m) \approx P(f)P(m|f)P(\ell|m)
\end{equation}

To mitigate the problem of zero probabilities of unseen name components, we use several smoothing techniques~\cite{chen1999,goodman2001}. Smoothing redistributes a portion of probability mass to not yet seen names. 

In case of LNRE distributions it is highly beneficial to have an estimate of unseen events for smoothing. LNRE models implemented in \textit{zipfR} package for R environment~\cite{zipfr} allow us, starting with name frequency distributions in the training set, to estimate the number of different names in a set of doubled size and consequently the number of names not appearing in the training set. As Table~\ref{tab:predict} (columns 1 and 2) shows, the \textit{Generalized Inverse Gauss-Poisson (GIGP)} model  implemented in \textit{zipfR} performs very well; the third column contains country-wide estimates for reference. 

\begin{table}[ht]
\centering
\caption{Prediction of the number of unique names}
\label{tab:predict}
\begin{tabular}{r R{1.7cm} R{1.7cm} R{2.5cm}}
\hline
Name &	$GIGP$ estimates &  Actual counts in $S$ & Country-wide  $GIGP$ estimates \\
\hline
$ f $              &   111,538 &   111,287 &    405,154\\
$ m $              &   155,635 &   155,726 &    462,738\\
$ \ell $           &   461,343 &   463,613 &    729,218\\
$ f \, \ell $      & 4,383,342 & 4,391,157 & 20,330,441\\
$ f \, m \, \ell $ & 9,088,527 & 9,087,716 & 65,867,708\\ 
\hline
\end{tabular}
\end{table}

\emph{Laplace smoothing} (models V and VI) is a simple additive smoothing method: pretend that every name $x$ occurs $\alpha > 0$ times more than it has been observed in the training set. Thus, the number of people with previously unseen name is $\alpha$. 
If $V$ is the set of unique names in $S_{train}$, then
\begin{equation}
\label{eq:laplace}
P_{L}(x) = 
\frac{C(x) + \alpha}{N + \alpha|V|}
\end{equation}

\emph{Good-Turing smoothing}~\cite{good1953} is a more gentle approach widely employed in language modeling (VII). The general idea behind the approach is to estimate the probability of all unseen names in the test set roughly equal to the total probability of names that appear only once in the training set, i.e. $\frac{N_1}{N}$. The counts of all other names are discounted accordingly:
\begin{equation}
C^*(x) = (C(x)+1)\frac{N_{C(x)+1}}{N_{C(x)}}.
\end{equation}
This results in the following probability estimates:
\begin{equation}
\label{eq:GT}
P_{GT}(x) = 
\begin{cases}
\frac{C^*(x)}{N}, & \mbox{if } C(x) > 0  \\
\frac{N_1}{N}\cdot \frac{1}{E}, & \mbox{if } C(x) = 0
\end{cases},
\end{equation}

\noindent where $E$ is a $GIGP$ estimate of hapaxes in $S$ based on $S_{train}$. Note that it implies we know the size of the test set  $S$ beforehand. 

One of the drawbacks of the Good-Turing smoothing is that it discounts probabilities uniformly in different frequency ranges. It leads often to severely distorted probabilities for high-frequency items. \emph{Katz smoothing}~\cite{Katz87} uses MLE for high-frequency names ($C(x) > 3$ in our experiment, model VIII) and Good-Turing smoothing for low-frequency ones.\footnote{Katz smoothing as described in the original work incorporates two approaches: 1)~combination of ML and GT estimates and 2)``backing off'' to lower-order n-grams in case of sparse data. In this study, we use only the former one.} 

\begin{equation}
\label{eq:KZ}
P_{K}(x) = 
\begin{cases}
P_{MLE}(x), & \mbox{if } C(x) > 3 \\
P_{GT}(x), & \mbox{if } 1 \leq C(x) \leq 3 \\
\frac{N_1}{N} \cdot \frac{1}{E}, & \mbox{if } C(x)=0 \\
\end{cases}
\end{equation}

Aiming at combining the simplicity of Laplace smoothing and the selectivity of Katz smoothing, we introduce \textit{pseudo-Laplace smoothing} with a small $\alpha>0$ (model IX):  

\begin{equation}
\label{eq:pseudoLaplace}
P^*_{PL}(x) = 
\begin{cases}
\frac{C(x)}{N + \alpha}, & \mbox{if } C(x) > 0 \\
\frac{\alpha}{N + \alpha}, & \mbox{if } C(x) = 0
\end{cases}
\end{equation}

The idea is quite simple: names present in the training set obtain probability close to the MLE, while unseen names get reasonable non-zero probabilities. In a strict mathematical sense, these are not probabilities, since they do not sum up to unity (and that is why we denote it $P^*$). Such probability-like scores are widely used in many practical applications, see for example ``stupid back-off'' introduced in~\cite{brants2007}.

\subsection{Experimental design}
\label{sec:experiment}

We conducted two experiments: 1) estimation of name popularity (that is, estimation of the number of people bearing each name) and 2) record linkage based solely on the name popularity estimates. In the first experiment we used a list of 13.4 million real-world persons represented by TINs and corresponding names compiled from the original dataset. In the second experiment, we performed record linkage on the original dataset of 20.6 million records.

\paragraph{Name popularity estimation.} Evaluation of models on samples with a large number of unique events is not an easy task. Evaluation results may diverge significantly on different test samples and depend on the size of test sample, particularly in low  frequencies ranges. For example, LNRE models are traditionally evaluated by looking at how well expected values generated by them fit empirical counts extracted from the same dataset used for parameter estimation~\cite{evert2004,baayen}. In this experiment we follow extrapolation setting for evaluation described in~\cite{evert2005}: the parameters of the model are estimated on a subset of the data used subsequently for testing. 
We randomly sampled a training set $S_{train}$ of 6.7 million names, which is $50\%$ of the whole dataset $S$.\footnote{We also performed experiments accounting for historical dimension: we ranked all persons with available year of birth by age and trained parameters on the `older' half of the population.
The results showed general decrease in quality, which supports the hypothesis of name popularity dynamics~\cite{kessler2012}. We do not cite the results here due to limited space.} 
We employ \emph{root-mean-square error} (RMSE) between the estimates and actual counts averaged over all names as evaluation measure. 
RMSE of the model $\mathcal{M}$ on the test set of people $S$ over the set of unique full names $V$ is defined as follows:\footnote{Note, that in this case $C(x)$ corresponds to the number of persons bearing name $x$ in $S$ (not in $S_{train}$ as in equations above). }
\begin{equation}
\sigma = \sqrt[2]{\frac{\sum_{x \in V}
(|S| \cdot P_\mathcal{M}(x) - C(x))^2}{|V|}}
\end{equation}
In order to have a better understanding of models' behavior and their applicability to different tasks and data volumes, we calculate $\sigma$ for the following name frequency buckets: $1$ (hapaxes), $2-5$, $6-20$, $21-100$, and $>100$ (very frequent names).

\paragraph{Record linkage.}
For the second task we calculate $P(|S| \cdot P_\mathcal{M}(x)<1)$, i.e. the probability that there is a single person with a given name $x$ in the population of size $|S|$ using estimates by different models $\mathcal{M}$. If the probability surpasses the threshold $t$, we link records with identical names. Note that all identical names are linked at once, whereby $q$ records with a given name trigger $\frac{q(q-1)}{2}$ linkages. The evaluation measure for the task are standard classification measures:  \textit{precision}~-- the fraction of linked records pairs that are correct, i.e. both refer to the same real-world person, and \textit{recall}~-- the fraction of correct links identified. As stated before, there are about 63.2 million pairs of identical names among 20.6 million occurrences, i.e. potential links between same-person records; 32\% of them are correct according to TINs. Taking into account these figures, linking all possible pairs results in $precision=32\%$ and $recall=100\%$. 

In contrast to the first experiment that presumably reflects a global distribution of names, the second experiment deals with a concrete database and its particular characteristics, e.g. the number of companies associated with a person. 

\section{Results}
\label{sec:results}

\subsection{Name count prediction}

\begin{table*}[ht]
\centering
\caption{Name models performance for full name triples}
\label{tab:results}
\begin{tabular}{l c r r r r r}
\hline
Model & Description & $\sigma_{1}$ & $\sigma_{2-5}$ & $\sigma_{6-20}$ & $\sigma_{20-100}$  & $\sigma_{>100}$ \\ \hline

I & Always 1 & 0.000 & 1.833 & 9.163 & 38.279 & 163.327 \\
II & $P_{MLE}(fml)$ & 1.000 & 1.611 & {\bf 3.061} &  {\bf 5.949} & {\bf 12.627}\\ 
III & $P_{MLE}(f) P_{MLE}(m) P_{MLE}(\ell)$ & 0.940 & 1.842 & 4.633 & 14.573 & 56.297 \\ 
IV & $P_{MLE}(f|m) P_{MLE}(m|\ell) P_{MLE}(\ell)$ & 0.897 & {\bf 1.608} & 3.165 & 6.639 & 16.925 \\ 
V & $P_{L}(f|m) P_{L}(m|\ell) P_{L}(\ell) \quad \alpha = 1$ & 0.999 & 2.720 & 9.779 & 36.277 & 137.747 \\ 
VI & $P_{L}(f|m) P_{L}(m|\ell) P_{L}(\ell) \quad \alpha = \frac{1}{|S_{train}|}$ & 0.897 & {\bf 1.608} & 3.165 & 6.639 & 16.925 \\ 
VII & $P_{GT}(f|m) P_{GT}(m|\ell) P_{GT}(\ell)$ & 0.900 & 1.622 & 3.171 & 6.644 & 16.931  \\ 
VIII & $P_{K}(f|m) P_{K}(m|\ell) P_{K}(\ell)$ & 0.901 & 1.614 & 3.165 & 6.639 & 16.925 \\ 
IX & $P^*_{PL}(f|m) P^*_{PL}(m|\ell) P^*_{PL}(\ell) \quad \alpha = 1$ & {\bf 0.885} & {\bf 1.608} & 3.165 & 6.639 & 16.925 \\ 
\hline
\end{tabular}
\end{table*}

\begin{table*}[ht]
\centering
\caption{Name models performance for first and last name doubles}
\label{tab:resultsFL}
\begin{tabular}{l c r r r r r }
\hline
Model & Description & $\sigma_{1}$ & $\sigma_{2-5}$ & $\sigma_{6-20}$ & $\sigma_{20-100}$  & $\sigma_{>100}$ \\ \hline

I & Always 1 & 0.000 & 1.956 & 9.676 & 44.018 & 356.760  \\ 
II & $P_{MLE}(f \ell)$ & 1.000 & 1.644 & 3.139 & {\bf 6.380} & 15.915 \\ 
III & $P_{MLE}(f) P_{MLE}(\ell)$ & 1.180 & 3.035 & 8.781 & 18.422 & 54.749 \\ 
V & $P_{L}(f|\ell) P_{L}(\ell) \quad \alpha = 1$ & 0.992 & 2.847 & 10.474 & 43.681 & 301.474  \\ 
VI & $P_{L}(f|\ell) P_{L}(\ell) \quad \alpha = \frac{1}{|S_{train}|}$ & 0.998 & 1.643 & 3.139 & {\bf 6.380} & 15.915 \\ 
VII & $P_{GT}(f|\ell) P_{GT}(\ell)$ & {\bf 0.665} & 1.487 & 3.359 & 6.443 & {\bf 15.912} \\ 
VIII & $P_{K}(f|\ell) P_{K}(\ell)$ & 0.727 & 1.659 & 3.638 & {\bf 6.380} & 15.915  \\ 
IX & $P^*_{PL}(f|\ell) P^*_{PL}(\ell) \quad \alpha = 1$ & 0.707 & {\bf 1.434} & {\bf 3.129} & {\bf 6.380} & 15.915  \\ 
\hline
\end{tabular}
\end{table*}

\begin{figure}
        \centering
        \begin{subfigure}{0.5\textwidth}
                \includegraphics[width=\textwidth]{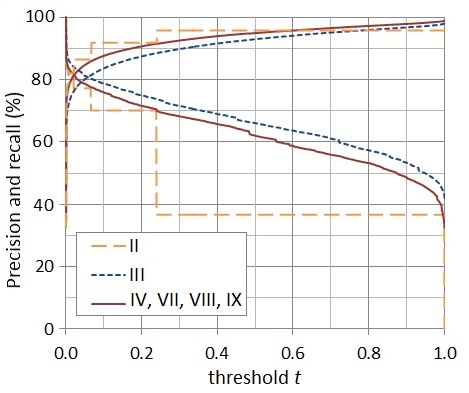}
                \caption{name triples (first, middle, last)}
                \label{fig:precrec_fio}
        \end{subfigure}%
        ~ 
        \begin{subfigure}{0.5\textwidth}
                \includegraphics[width=\textwidth]{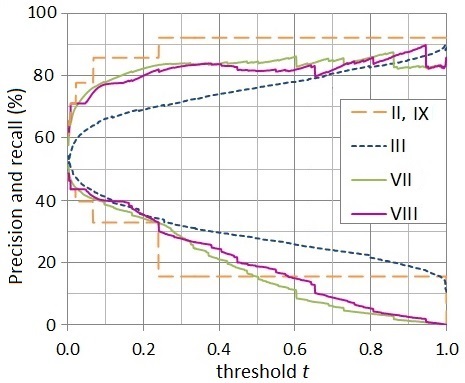}
                \caption{name doubles (first, last)}
                \label{fig:precrec_fi}
        \end{subfigure}
        \caption{Record linkage evaluation results: precision (upper curves) and recall (lower curves) of various name count prediction methods depending on the threshold value $t$}
        \label{fig:linkage}
\end{figure}

Table \ref{tab:results} summarizes evaluation results for nine name popularity prediction models. The first model (I) is a na\"{i}ve ``always 1'' baseline that assumes all names are unique. Obviously, the model performs ideally on hapaxes.  
MLE model for full name triples (II) demonstrates the best prediction results in higher frequency ranges. The product of individual probabilities for first, middle and last names (III) performs slightly better on hapaxes, but substantially underestimates the probability of more frequent names. We investigated different dependencies between full name constituents, and combination in the model IV performed best. As one can see, conditional probabilities considerably improve over model III that assumes independence of name constituents. 

The next five models incorporate smoothing. Add-1 smoothing (V) is too aggressive in case of LNRE distributions, as the evaluation results show. A more delicate Laplace smoothing with $\alpha = \frac{1}{|S_{train}|}$ (VI) delivers better results that are equal to model IV's ones. Good-Turing and Katz methods with $GIGP$ estimates (VII and VIII, respectively) perform slightly worse, but comparably to other models with smoothing. Our method (IX) performs best in the low-frequency range and equally well as models IV and VI in higher-frequency areas. 

Table \ref{tab:resultsFL} summarizes performance of the same models for first-last name doubles (model IV applied to name doubles coincides with model III). Although the general trend is the same as in case of name triples, the results contain some peculiarities. Good-Turing model (VII) is the best in predicting hapaxes and most frequent names. The latter fact is somewhat unexpected, and we will study this outcome in depth in our future work. The proposed pseudo-Laplace method performs best in the middle frequency range (2--100). 

\subsection{Record linkage}

Results of the record linkage experiment 
are presented in Figures~\ref{fig:precrec_fio} (name triples) and~\ref{fig:precrec_fi} (name doubles). The threshold $t$ governs the linkage process: the higher the threshold the less name mentions are linked. One can imagine the process of gradual data linkage going from right to left, from higher to lower $t$ values. Stepped curves of the MLE models are due to the fact that at some $t$ values a large number of links is established at a time.
In the case of full name triples (Figure~\ref{fig:precrec_fio}) all `advanced' methods deliver almost identical results. The simplest MLE method for full names works well when we favor precision over recall. Threshold $t=0.2$ delivers precision of about 90\% and recall above 70\%. In the case of first and last name doubles, the task of record linkage in such a sizable dataset based solely on name popularity estimates is much less effective (see Figure~\ref{fig:precrec_fi}).

\section{Conclusion and future work}
\label{sec:conclusion}

In our experiments we make use of a large name dataset with unique identifiers that contains names of approximately one tenth of the Russian population. 
We conducted a series of experiments with different name popularity prediction models built upon the name dataset. We thoroughly evaluated several models, including well-known smoothing approaches and proposed a new simple yet effective method for adjusting probability estimates accounting for unseen events. Results show that the considered methods behave differently depending on the frequency range of names to be estimated, the name structure (full name triples vs. first and last name doubles), and the population size for which the prediction is made. These experimental results can serve as guidelines for choosing the most suitable method for a specific task and available data. 

Furthermore, we conducted a record linkage experiment in the large database based solely on name popularity estimates. The outcomes suggest that name popularity estimates are a valuable signal for personal name matching. Results show that all methods using smoothing perform almost identically and the simplest method based on maximum likelihood estimates can be a good choice, when precision is more important than recall. However, these results reflect the peculiarities of a specific database and serve merely as an illustration of feasibility of the approach. 

Proposed statistical techniques can incorporate other components along names such as location, gender, age and so on. In case of our dataset, locations associated with a person can be derived either from TIN -- it encodes the federal district, where the TIN was issued, -- or from the legal address of the associated company. An record linkage experiment accounting for location ($loc$) in the form $P(x) = P(loc) \cdot P^*_{PL}(\ell|loc) \cdot P^*_{PL}(m|\ell) \cdot P^*_{PL}(f|m)$ achieved precision 95\% and recall 83\% on the dataset. 

The proposed methods are applied to identical name strings and do not account for misspellings, OCR errors, spelling and transliteration variants. An interesting direction for future research could be combination of name popularity estimates and string similarity measure traditionally used in record linkage tasks. 

In future work we plan to incorporate other sources of name popularity information such as phone books, open electoral registers, and social network sites and to compare results obtained using different datasets. It is also interesting to juxtapose name popularity distributions in different countries and cultures. 

\subsubsection{Acknowledgements.} We thank Kontur for preparing the dataset and granting access to it for research. We are very grateful to Leonid Boytsov, Julia Efremova, James Lu, Boris Novikov, Guillaume Obozinski, Julia Stoyanovich, and Yana Volkovich for reading the paper draft and making valuable comments and suggestions. 

\bibliographystyle{splncs04}
\bibliography{all}

\end{document}